\documentclass[conference]{IEEEtran}
\IEEEoverridecommandlockouts
\usepackage{cite}
\usepackage{amsmath,amssymb,amsfonts}
\usepackage{algorithmic}
\usepackage{graphicx}
\usepackage{textcomp}
\usepackage[normalem]{ulem}
\usepackage{xcolor}
\usepackage{epstopdf}
\usepackage{url}
\usepackage{bm}
\usepackage{bbm}
\usepackage{upgreek}
\usepackage[super]{nth}
\usepackage[margin=0.75in]{geometry}
\usepackage{cancel}

\newcommand{\Hnull}{\mathcal{H}_0}
\newcommand{\Halt}{\mathcal{H}_1}

\newcommand{\Honull}{\mathcal{{D}}_0}
\newcommand{\Hoalt}{\mathcal{{D}}_1}

\begin{document}


\title{Machine Learning and Location Verification in Vehicle Networks\\
}

\title{{Machine Learning and Location Verification in Vehicular Networks}}
\author{
\IEEEauthorblockN{Ullah Ihsan$^1$, Robert Malaney$^1$ and Shihao Yan$^2$}
\IEEEauthorblockA{$^1$School of Electrical Engineering  \& Telecommunications,
The University of New South Wales,
Sydney, NSW 2052, Australia \\
$^2$School of Engineering,
Macquarie University,
Sydney, NSW 2109, Australia}}

\maketitle

\begin{abstract}

Location information  will play a very important role in emerging wireless networks such as Intelligent Transportation Systems, 5G, and the Internet of Things. However, wrong location information can  result in poor network outcomes. It is therefore critical to verify all  location information before further utilization in any network operation. In recent years, a number of information-theoretic Location Verification Systems (LVSs) have been formulated in attempts to optimally verify the location information supplied by network users. Such LVSs, however, are somewhat limited since they rely on knowledge of a number of  channel parameters for their operation. To overcome such limitations, in this work we introduce a Machine Learning based LVS (ML-LVS). This new form of LVS can adapt itself to changing environments without knowing the channel parameters.
Here, for the first time, we use real-world data to show how our ML-LVS can outperform information-theoretic LVSs. We demonstrate this improved performance within the context of vehicular networks using Received Signal Strength (RSS) measurements at multiple verifying base stations. We also demonstrate the validity of the ML-LVS even in scenarios where a sophisticated adversary optimizes her attack location.

\end{abstract}


\section{Introduction}\label{introduction}

Vehicular Ad-hoc Networks (VANETs) are a particular type of Intelligent Transportation System (ITS) which utilize communications to assist with various traffic problems. VANETs can function based on vehicle-to-vehicle communication and/or vehicle-to-Road Side Unit (RSU) communication\cite{hartenstein2008tutorial}. RSUs are fixed base stations installed at certain locations with an aim to assist VANETs with their operations. An RSU (or a trusted vehicle  whose location is \emph{a priori} verified), can also function as a Processing Center (PC). The PC processes the communication data before issuing instructions to the vehicles under its coverage area.

Location information of vehicles is a key ingredient for VANETs. The vehicles usually obtain their location information through Global Navigation Satellite System (GNSS) and/or Global Positioning System (GPS), and report this information to the PC for use in subsequent network operations. A possibility exists where the supplied location information from the vehicle has errors in it. This may be due to some faulty hardware used in recording/forwarding the location information, or it may be due to a vehicle falsifying its location information (in order to have advantage over nearby vehicles or to simply disrupt the network). If the location information supplied by the vehicle is not verified, and the location error goes unnoticed, this may result in poor network outcomes such as traffic queues, traffic congestion, or poor tolling. In extreme cases, a lack of position verification may lead to catastrophic situations such as vehicle collisions.

 In recent years, a number of 
 Location Verification Systems (LVSs)\cite{yan2013location,sheet2016location,yan2016location,monteiro2016information,rob2016location,malandrino2013vip,jaballah2013secure,kim2017efficient,caparra2017optimization,vaas2018get} have been devised to validate the vehicle's supplied location information. These LVSs in general make use of the numerous physical layer properties of the signal (transmitted by the vehicle and measured at the verifying base stations) to verify the vehicle's reported location information. The physical layer properties include Received Signal Strength (RSS), Time of Arrival (ToA) of the signal, and Angle of Arrival (AoA) of the signal. However, all 
 LVSs have a serious limitation in their operation - they normally operate efficiently only for the channel conditions assumed at the time of their design\cite{yan2013location}. That is, they normally only function well under the assumption that all \emph{a priori}  channel information provided to them remains accurate. Further, they are only able to efficiently address the threat-model scenarios they have been specifically designed for\cite{yu2013detecting}. Such limitations make their  real-world deployment suspect.

Machine Learning (ML) is an important technology which is now impacting many applications e.g.,\cite{abadi2016tensorflow,wan2014deep,rosten2006machine,hinton2012deep,saad,matvejka2016analysis,ze2013statistical},
and it is possible that inclusion of ML techniques may help resolve some of the LVS limitations mentioned above. Indeed, this has been shown to be the case in
 theoretical simulations of LVSs in the context of ToA schemes\cite{ihsan2019artificial}, and in theoretical simulations of `in-region' location verification\cite{brighente2018location}. What remains to be determined is whether these advances hold up under conditions where real-world data is input to the ML-LVS. In this work, which represents the first experimental deployment of any ML-LVS, we answer this question in the affirmative. We summarize below our main contributions.

\begin{itemize}
\item {We carry out for the first time an ML-LVS analysis based on real-world data, namely RSS measurements}.
\item {We show that our ML-LVS outperforms an  information-theoretic LVS when a malicious vehicle sets its claimed (untrue) location at some random location.}
\item {We also show that unlike the information theoretic LVS, the ML-LVS still performs efficiently even when the malicious vehicle formally minimizes spoofing detection by optimizing its claimed (untrue) location.}
\end{itemize}

The remainder of this paper is organized as follows. Section \ref{SM} details the system model. Section \ref{PA} presents the performance analysis using information theory and ML techniques. Section \ref{NR} provides numerical results and future prospects, and Section \ref{Conclusion} concludes the paper.

\section{System Model}\label{SM}

We consider the following system model in our work:
\begin{enumerate}
\item {The true location of a legitimate or malicious vehicle is denoted by $\textbf{x}_t=[x_t,y_t]$}.
\item {We refer to the reported location from a legitimate or malicious vehicle as the \emph{claimed location}, which is denoted by $\textbf{x}_c=[x_c,y_c]$. The claimed location for a legitimate vehicle is exactly the same as its true location. On the other hand, a malicious vehicle spoofs its location, \emph{i.e.}, its claimed which is not the same as its true location.}
\item {For a malicious vehicle $||\textbf{x}_c \mathrm{-} \textbf{x}_t || \geq r$, where $r$ is an \emph{a priori} distance representing the minimum distance between its claimed and true locations.}
\item {The framework consists of $N$ RSUs as verifying base stations, with publicly known true locations. All RSUs are in the transmission range of the vehicles (whose claimed locations have to be verified). The true location of the \textit{i}-th RSU is $\textbf{x}_i=[x_i,y_i]$ where $i=1,2,...,N$.}
\item {We choose one of the RSUs as PC. The PC accumulates its own RSS measurements with the measurements collected by other RSUs for further processing. The PC decides on the integrity of a vehicle's claimed location.}
\item {Under the null hypothesis $\mathcal{H}_o$, the vehicle is legitimate, \emph{i.e.}, we have}
\begin{align}
\mathcal{H}_o : \textbf{x}_c=\textbf{x}_t.
\end{align}
\item {Under the alternative hypothesis $\mathcal{H}_1$, the vehicle is malicious, \emph{i.e.}, we have}
\begin{align}
\mathcal{H}_1 : \textbf{x}_c\neq \textbf{x}_t.
\end{align}
\end{enumerate}
Based on a log-normal pathloss model, under $\mathcal{H}_o$, the RSS (all RSS in dBm) measured by the \textit{i}-th RSU from a legitimate vehicle, $y_i$, is given by
\begin{align}\label{eq3}
y_i=u_i+w_i,\ \ \ \ \ \ \ \ \ \ \ \ i=1,2,\dots ,N,\
\end{align} where $w_i$ is a zero mean normal random variable with variance $\sigma^2_{T}$ representing the channel noise, and $u_i$ is the mean RSS at \textit{i}-th RSU. This latter quantity is given by\\
\begin{align}\label{distance_time}
u_i= p_{d_o}\mathrm{-}10\,\gamma\,\log_{10}\Big(\frac{d_i^c}{d_o}\Big),
\end{align}
where $p_{d_o}$ is a reference RSS at a reference distance $d_o$, $\gamma$ is the path loss exponent, and $d_i^c$ is the  distance of a legitimate vehicle's true location to the \textit{i}-th RSU, given by
\begin{align*}
d_i^c=\sqrt{{(x_c-x_i)}^2+{(y_c-y_i)}^2}.
\end{align*}
The measurements made by the \textit{N} RSUs are independent of each other. Under $\mathcal{H}_o$, they collectively form a vector $\textbf{y}={[y}_1,\ y_2,\dots ,{y_N]}^T$. Based on  \eqref{eq3} the vector $\textbf{y}$ follows a multi-variate normal distribution given as
\begin{align}
f(\textbf{y}| \mathcal{H}_o) \sim \mathcal{N} (\textbf{u}, \Upsigma),
\end{align}
where $\textbf{u} = {[u}_1,\ u_2,\dots ,{u_N]}^T$ is the mean RSS vector under $\mathcal{H}_o$, and $\Upsigma = \sigma_T^2 \textbf{I}_N$ is the covariance matrix with $\textbf{I}$ as the identity matrix.
\\\\
\noindent Under $\mathcal{H}_1$, a malicious vehicle spoofs its claimed location. It reports its claimed location to be at a minimum distance $r$ away from his true location. As an example scenario -  we can think of the  malicious vehicle pretending to be on the road while it actually is placed off in a nearby street. The RSS value measured by the \textit{i}-th RSU from a malicious vehicle, $y_i$, is given by
\begin{align}\label{eq6}
y_i=v_i+w_i,\ \ \ \ \ \ \ \ \ \ \ \ i=1,2,\dots ,N,\
\end{align}
where $v_i$ is given by
\begin{align}\label{distance_time2}
v_i= p_{d_o}\mathrm{-}10\,\gamma\,\log_{10}\Big(\frac{d_i^t}{d_o}\Big),
\end{align}
and $d^t_i$ is the  distance of its true location to the \textit{i}-th RSU, given by
\begin{align*}
d^t_i=\sqrt{{(x_t-x_i)}^2+{(y_t-y_i)}^2}.
\end{align*}
The measurements made by \textit{N} RSUs are independent of each other. Under $\mathcal{H}_1$, they collectively form a vector $\textbf{y}={[y}_1,\ y_2,\dots ,{y_N]}^T$. From  \eqref{eq6}, vector $\textbf{y}$ follows a multi-variate normal distribution given as
\begin{align}
f(\textbf{y}| \mathcal{H}_1) \sim \mathcal{N} (\textbf{v}, \Upsigma),
\end{align}
where $\textbf{v} = {[v}_1,\ v_2,\dots ,{v_N]}^T$ is the mean RSS vector under $\mathcal{H}_1$.

\section{Performance Analysis}\label{PA}
The outcome of an LVS is a binary result i.e. legitimate or malicious. This is different from a localization system where the output is an estimated location. We measure the performance of our LVS using two methodologies; through information theoretic analysis similar to \cite{yan2014timing} and, through the newly designed ML-LVS method which makes use of machine-learning techniques. In both the cases, a Bayes average cost function is chosen as the performance metric for LVS in terms of `Total Error'. The Total Error is given by
\begin{align} \label{total_error}
\mathrm{\xi} = p(\mathcal{H}_o)\upalpha + p(\mathcal{H}_1)(1-\upbeta),
\end{align}
where $p(\mathcal{H}_o)$ and $p(\mathcal{H}_1)$ are the \emph{a priori} probabilities of occurrences of $\mathcal{H}_o$ (i.e. legitimate vehicle) and $\mathcal{H}_1$ (i.e. malicious vehicle), respectively. In this work, we assume the legitimate and the malicious vehicles in equal proportions so both $p(\mathcal{H}_o)$ and $p(\mathcal{H}_1)$ are equal to 0.5. $\upalpha$ represents the False Positive Rate (the rate of legitimate vehicles being detected incorrectly) and $\upbeta$ represents the Detection Rate (the rate of malicious vehicles being detected correctly). Equation \eqref{total_error} therefore takes the form
\begin{align}\label{totalerror}
\mathrm{\xi } = \mathrm{0.5} \upalpha \mathrm{\:+\:0.5}\left(\mathrm{1-}\upbeta \right).
\end{align}
\subsection{Information-theoretic LVS}
We will refer to the information-theoretic analysis as the Likelihood Ratio Test (LRT) method from now on. The LRT method requires some parameters and channel information to be available in advance. This information includes the pathloss exponent $\gamma$, the mean RSS vectors as highlighted in the system model, and the LRT decision threshold $\ell$,
It has been proven elsewhere that the LRT method achieves the optimum detection results for a given false positive rate\cite{neyman1933ix}. This leads to the conclusion that the LRT minimizes the Total Error and maximizes the mutual information between input and output of the LVS\cite{yan2014optimal}. We follow decision rule given below for the LRT method
\indent
\begin{align}\label{LRT}
\Lambda \left(\textbf{y}\right)\triangleq \frac{p(\textbf{y}|\mathcal{H}_1)}{p(\textbf{y}|\mathcal{H}_o)}\ \genfrac{}{}{0pt}{}{\genfrac{}{}{0pt}{}{\Hoalt}{\ge }}{\genfrac{}{}{0pt}{}{<}{\Honull}} \ell,
\end{align}
where $\Lambda \left(\textbf{y}\right)$ is the likelihood ratio, and $\Hoalt$ and $\Honull$ are the binary decision values (\emph{i.e.}, whether the vehicle is legitimate or malicious), while $p(\textbf{y}|\mathcal{H}_o)$, and $p(\textbf{y}|\mathcal{H}_1)$ are given by
\begin{align}\label{new_1}
p(\textbf{y}|\mathcal{H}_o)=\frac{1}{\sqrt[k]{2\pi}\sqrt{|\Upsigma|}}e^{\mathrm{-}\frac{1}{2}(\textbf{y}\mathrm{-}\textbf{u})\Upsigma^{-1}  (\textbf{y}\mathrm{-}\textbf{u})},
\end{align}
\begin{align}\label{new_2}
p(\textbf{y}|\mathcal{H}_1)=\frac{1}{\sqrt[k]{2\pi}\sqrt{|\Upsigma|}}e^{\mathrm{-}\frac{1}{2}(\textbf{y}\mathrm{-}\textbf{v})\Upsigma^{-1}  (\textbf{y}\mathrm{-}\textbf{v})},
\end{align} where $|\Upsigma|$ is determinant of $\Upsigma$. The decision rule given in \eqref{LRT} can be reformulated as
\indent
\begin{align} \label{LRT2}
\Lambda \left(\textbf{y}\right)\triangleq \frac{e^{\mathrm{-}\frac{1}{2}(\textbf{y}\mathrm{-}\textbf{v})\Upsigma^{-1}  (\textbf{y}\mathrm{-}\textbf{v})}}{e^{\mathrm{-}\frac{1}{2}(\textbf{y}\mathrm{-}\textbf{u})\Upsigma^{-1}  (\textbf{y}\mathrm{-}\textbf{u})}}\ \genfrac{}{}{0pt}{}{\genfrac{}{}{0pt}{}{\Hoalt}{\ge }}{\genfrac{}{}{0pt}{}{<}{\Honull}} \ell.
\end{align}
We assume that the malicious vehicle optimizes its claimed location.  That is, through an  optimization strategy, it  minimizes its probability of being detected by the LVS. We assume in this work that the malicious vehicle's optimum claimed location is constrained to be within  the transmission range of the RSUs. To optimize its claimed location under such a constraint, the malicious vehicle minimizes the KL divergence between $f(\textbf{y}| \mathcal{H}_1)$ to $f(\textbf{y}| \mathcal{H}_o)$\cite{eguchi2006interpreting}. This divergence is as given below
\begin{equation}\label{KLdistance}
\begin{split}
D_{KL}(f(\textbf{y}|\Halt)|| f(\textbf{y}|\Hnull)) &= \int_{-\infty}^{\infty} f(\textbf{y}|\Halt) \ln {\frac{f(\textbf{y}|\Halt)}{f(\textbf{y}|\Hnull)}} d{\textbf{y}},\\
&= \frac{1}{2}(\textbf{v} \mathrm{-} \textbf{u})^T \, \Upsigma^{-1} (\textbf{v} \mathrm{-} \textbf{u}).
\end{split}
\end{equation}
Then, the optimal claimed location $\textbf{x}_c^{\ast}$ for the malicious vehicle can be obtained through
\begin{equation}\label{optimal xc}
\begin{split}
\textbf{x}_c^{\ast} &= \underset{||\textbf{x}_t \mathrm{-} \textbf{x}_c||\geq r}{\mathrm{argmin}}D_{KL}(f(\textbf{y}|\Halt)|| f(\textbf{y}|\Hnull)).\\
\end{split}
\end{equation}

\subsection{ML-LVS}

This section highlights the novel approach used to design a classification framework for the verification of a vehicle's claimed location through supervised ML techniques. Feed-forward neural networks are well known for their performance in classification problems. We use a multi-layer feed-forward neural network for the binary classification of a vehicle as either legitimate or malicious.

The framework considers $\textbf{y}$ (the RSS observation vector measured in the field) and the vehicle's claimed location   as inputs. Based on a series of trials with changing architectures for the ML-LVS, we decided upon a framework that has 
the raw inputs (RSS, claimed locations, and RSUs locations), a 10-neuron  hidden layer, and a 1-neuron binary output layer. We also experimented with  different transfer functions in various layers of the ML-LVS. The results shown in the next section adopted the hyperbolic tangent-sigmoid transfer function in the hidden layer and the linear transfer function in the output layer. The ML-LVS utilized the Levenberg-Marquardt as its backpropagation algorithm.

\section{Numerical Results}\label{NR}

\begin{figure}[h]
\centering
\includegraphics[width=0.40\textwidth]{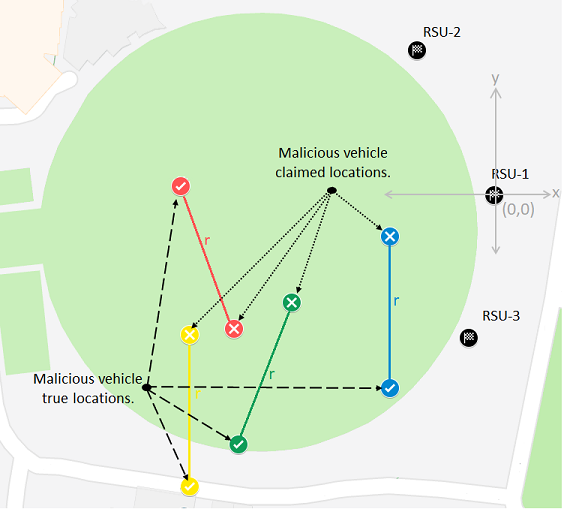}
\caption{Malicious vehicles fake their locations to launch a location-spoofing attack. They report their claimed locations $r$ meters away from their respective true locations. This figure only shows a sample of the malicious vehicles' true locations, their simulated random claimed locations at $r$ meters, and the true locations for the RSUs. The value of $r$ in this figure is 50 meters.\label{oval}}
\end{figure}

RSS measurements from the vehicles were collected in  a 150 X 150 meters area by 3 RSUs (an area that mimics a wide cross section of 2 highways). 3 devices were used as 3 RSUs to independently measure the RSS from the vehicles in the field at a frequency of 1 Hz simultaneously, \emph{i.e.}, one RSS measurement per second per RSU. The origin of the area is set to the location of RSU-1 as shown in Fig.~\ref{oval}.  Moving Wi-Fi modems  with a single antenna and an attached GPS (used to record the vehicle's location at a frequency of 1 Hz) was used to mimic  slow-moving  vehicles. The GPS locations of these `vehicles' are reported to the RSUs every second. The RSS measurements by individual RSUs and the vehicles' GPS locations were combined with the help of time stamps (available with both the measured RSS and the vehicles' GPS locations).

The pathloss exponent $\gamma$ is required for the LRT, and is determined directly from the field measurements via a linear fit of the measured RSS values against the logarithm of the distance to a RSU.   $\textbf{u}$ and $\textbf{v}$ are calculated using \eqref{distance_time} and \eqref{distance_time2} under the corresponding hypothesis. $\sigma_T$ is calculated using the mean RSS vector and the RSS measurements (made by each RSU).

The RSS measurements data is randomized and equally divided it into two halves with one half representing the legitimate vehicles and the other half representing the malicious vehicles. To launch a location-spoofing attack, the malicious vehicles spoof their locations by a minimum distance of $r$ meters away from their true locations. Random claimed locations for the malicious vehicles are simulated by taking into account the distance constraint $r$. Fig.~\ref{oval} highlights true and simulated random claimed locations for a sample of the malicious vehicles.

\begin{figure}[h]
\centering
\includegraphics[width=0.50\textwidth]{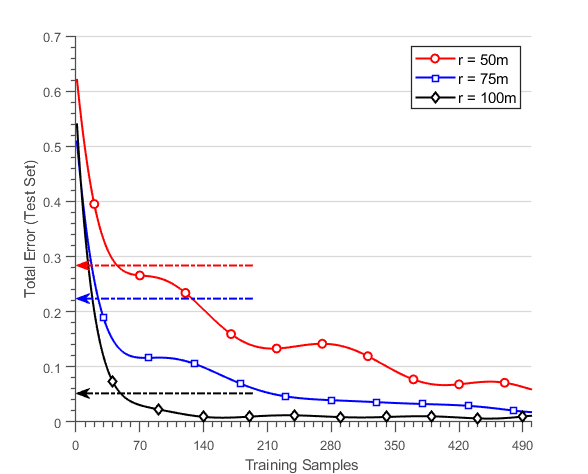}
\caption{A comparison study where an ML-LVS outperforms an LRT based LVS.  The ML-LVS with no \emph{a priori} channel information achieves a final Total Error (indicated by solid lines) of 0.01, 0.02, and 0.06, for $r$ equal to 100m, 75m, and 50m, respectively, for the data in the test set. On the other hand, the LRT based LVS with \emph{a priori} channel  information achieves a Total Error (indicated by dashed arrows) of 0.05, 0.22, and 0.29, for $r$ 100m, 75m, and 50m, respectively, for the data in the test set. Note, in these calculations, the malicious vehicles do not optimize their claimed locations.\label{Final}}
\end{figure}

We now present some numerical results based on our analysis from the LRT and ML-LVS.  In Fig.~\ref{Final}, we assume that the malicious vehicles \emph{randomly} forge their claimed locations at a minimum distance $r$ away from their true locations and within the transmission range of the RSUs. The Total Error is plotted against the number of training data used. For the LRT based LVS, we calculate the Total Error, the false positive rate, and the detection rate  under different values of $r$ using \eqref{totalerror} and \eqref{LRT2}. The Total Error for $r$ equal to 100m, 75m and 50m, is 0.05, 0.22, and 0.29, respectively (different colored-dashed arrows).

The data considered for the LRT based LVS in Fig.~\ref{Final} is also considered for the ML-LVS. Unlike the LRT method where the LVS requires \emph{a priori} information for the channel parameters, the ML-LVS only uses the measured RSS (at the RSUs) and the vehicles' reported claimed locations. This data which has genuine and malicious vehicles in equal proportions is randomized and divided into two data sets; a training set with 80\% of the entire data, and a test set with the remaining 20\% of the data. The training set
also has data labels (genuine or malicious). These data labels indicate whether particular training sample represents a legitimate or malicious vehicle. Use of such data is required to set the weights and biases for the ML-LVS in the training phase. On the other hand, the data in the test set has no such labels which means that we have no \emph{a priori} information if a particular sample belongs to a legitimate or a malicious vehicle. Once trained, the ML-LVS can be used to test the data in the test set for classification of the vehicles.

\begin{figure}[h]
\centering
\includegraphics[width=0.50\textwidth]{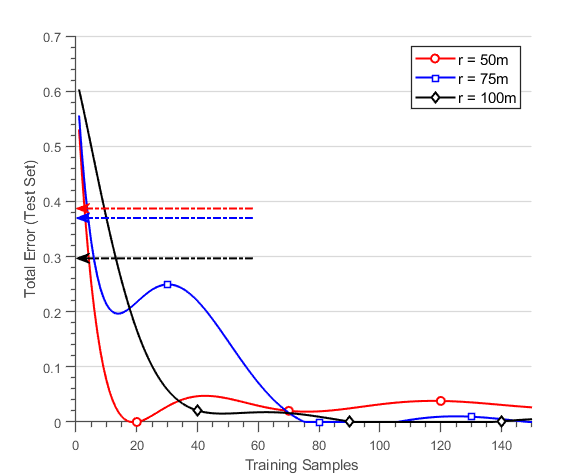}
\caption{A comparison study of the LRT based LVS with ML-LVS as in Fig.~\ref{Final} except the malicious vehicles now optimize their claimed locations}.\label{opt_plot}
\end{figure}

In the training phase in Fig.~\ref{Final}, we supply the ML-LVS with training samples from the training data at a rate of one random training sample per unit time and plot the Total Error for the test set after each unit time.  The ML-LVS's backpropagation algorithm terminates the training phase once a threshold for any of its internally set parameters is met. We observe that in most  cases the `maximum validation failures' parameter of the backpropagation algorithm (the maximum number of sequential iterations in which the ML-LVS's performance fails to improve) is reached, and this terminates the training phase. We set this parameter to 6.  This trained ML-LVS is then used to classify vehicles in the test set as either legitimate or malicious. This procedure is repeated for each value of the training data shown in Fig.~\ref{Final}.
As shown in Fig.~\ref{Final}, as expected, the Total Error for the test set improves as the training continues. The final Total Error for the test set (after 500 training samples) using the ML-LVS  for $r$ equal to 100m, 75m and 50m, is 0.01, 0.02, and 0.06, respectively. It is evident from Fig.~\ref{Final} that the ML-LVS with no \emph{a priori} channel  information has much-improved performance relative to the LRT based LVS.

We now assume that the malicious vehicles can  overhear the communication between the legitimate vehicles and the RSUs. The malicious vehicles use this information to best optimize their claimed locations ($\textbf{x}_c = \textbf{x}_c^{\ast}$) prior launching a location-spoofing attack. That is, they set their claimed location using \eqref{KLdistance} so as to minimize their probability of being marked malicious by the LVS.

In Fig.~\ref{opt_plot} we compare the performances of the ML-LVS and the LRT based LVS. We  see again that the ML-LVS still outperforms the LRT based LVS. However, we notice a rather counter-intuitive finding where, compared to Fig.~\ref{Final}, the Total Error for the ML-LVS improves much faster. This counter intuitive finding is as a result of the geometry of the RSUs in this specific experiment. This geometry leads to a clustering in the malicious vehicles' claimed location settings. In general (\emph{i.e.} more general RSU geometries), if the malicious vehicles' optimize their claimed locations, the Total Error for the ML-LVS is expected to take longer to reach its asymptotic  value.

In future work we plan to integrate Support Vector Machines (SVM) into the designed neural-network framework of our ML-LVS. We also plan to deploy this modified ML-LVS  in more complex channel fading environments such as those possessing Rician fading channels.
These additional studies are likely to provide for even more performance gains in ML-LVSs relative to  LRT based LVSs.

\section{Conclusion}\label{Conclusion}

Information-theoretic LVS frameworks, due to their operating limitations, are not practical in many real-world scenarios. To address this gap, we have proposed  the use of a ML approach to location verification. This new approach is particulary useful since unlike an information-theoretic LVS, a ML-LVS does not require \emph{a priori} information on the channel parameters. Additionally, a ML-LVS can adapt itself to any changing channel conditions.

Using real-world RSS data, we have shown for the first time how a deployed ML-LVS outperforms  state-of-the-art information-theoretic LVS. Further, we have shown how this result holds even when the adversary  optimizes its attack location. Future work in this area will help us develop a fully robust state-of-the-art artificially intelligent LVS, an LVS which will be wholly practical in terms of its location verification performance in a wide range of future wireless networks beyond the networks we have studied here.

We believe the novel approach for enhancing the performance of real-world LVSs that we have developed here potentially forms the foundation for all future works in the important area of wireless location verification.

\section{Acknowledgment}
The authors acknowledge support by the University of New South Wales, Australia, and Macquarie University, Australia.
Ullah Ihsan acknowledges financial support from the Australian Government through its Research Training Program.

\bibliographystyle{IEEEtran}
\bibliography{ICCC}

\end{document}